\documentclass[twoside]{sfm}

\input{sf.def}

\begin{document}

\def\llm{{\sc LLmodels}}
\def\atl{{\sc ATLAS9}}
\def\aatl{{\sc ATLAS12}}
\def\starsp{{\sc STARSP}}
\def\aur{$\Theta$~Aur}
\def\logg{\log g}
\def\tauros{\tau_{\rm Ross}}
\def\kms{km\,s$^{-1}$}
\def\bz{$\langle B_{\rm z} \rangle$}
\def\degr{^\circ}
% journals
\def\aaps{A\&AS}
\def\aap{A\&A}
\def\apjs{ApJS}
\def\apj{ApJ}
\def\rmxaa{Rev. Mexicana Astron. Astrofis.}
\def\mnras{MNRAS}
\def\actaa{Acta Astron.}
\newcommand{\Tef}{T$_{\rm eff}$~}
\newcommand{\Vt}{$V_t$}
\newcommand{\CC}{$^{12}$C/$^{13}$C~}
\newcommand{\CDC}{$^{12}$C/$^{13}$C~}

\pagebreak

\thispagestyle{titlehead}

\setcounter{section}{0}
\setcounter{figure}{0}
\setcounter{table}{0}

%%%%%%%%%%%%%%%%%%%%%%%%%%%
% !!!
\markboth{North}{Multiplicity of A-type stars}

\titl{Multiplicity of A-type and related stars}{North P.}
{Laboratory of Astrophysics, Ecole Polytechnique F\'ed\'erale de Lausanne (EPFL), Switzerland,
email: {\tt pierre.north@epfl.ch}}

\abstre{
The origin of chemically peculiar stars remains enigmatic, especially
regarding their frequency among their "normal" peers. In addition to
magnetic fields and rotation, multiplicity may shed light on the question.
We mention the main surveys of the three kinds performed so far of intermediate
mass stars, either normal or chemically peculiar, magnetic or not:
imaging, spectroscopic, and photometric. We also consider the multiplicity
of red giant stars, since many of them are descendants of A-type stars,
through Mermilliod's radial velocity monitoring of open cluster members.
We briefly review the orbital properties of binary systems hosting chemically peculiar stars.
Some specific objects of special interest are mentioned as deserving further study.
Finally, we recall that some binary systems composed of A-type stars are progenitors of
Type Ia supernovae, and evoke the potentialities of future surveys such as Gaia.
}

\baselineskip 12pt

\section{Introduction and some generalities}

"Multiplicity" refers here to a star accompanied by one or more stellar companions,
tied by gravitation or showing at least a common spatial motion. By "A-type and related
stars", we mean an initial mass range between about 1.3 and 8 M$_\odot$. This range is physically
defined by the conditions that the star hosts a convective core and a radiative envelope
 while on the main sequence, and ends its life as a white dwarf. It also includes most
chemically peculiar stars, because neither surface convection nor wind is strong enough
to prevent abundance anomalies to build up through radiative diffusion. Tokovinin's catalogue
of multiple stars \cite{T97} is worth mentioning here.
Duch\^ene \& Kraus \cite{DK13} summarize in their review the following essential facts
emerging from the study of multiple stars:

-- The multiplicity properties are a smooth function of the mass of the primary.

-- The frequency and properties of multiple stars are generally defined already
at the pre-main sequence phase. They do not evolve much afterwards.

-- The mass ratio ($q$) distribution cannot be explained by random pairing of two objects
taken from the Initial Mass Function (IMF), and eccentricity does not follow the so-called
thermal distribution $f(e)=2e$.

-- There is no clear upper limit to the binary separation, which may reach 1 pc or so.

In addition, there is a monotonous increase of both the frequency of multiple systems (MF)
and of the companion frequency (CF, or average number of companions to a given primary) with
the mass of the primary. The mass ratio $q=M_2/M_1$ follows a distribution that is generally
represented by a power law $f(q)\propto q^\gamma$, where $\gamma$ is close to zero or slightly
negative for intermediate mass stars, while it may increase up to $\sim 4$ from solar mass
stars down to the smallest stellar masses.

\section{Astrometric, radial velocity and photometric surveys}

For the normal stars, the main recent and ongoing astrometric survey is the VAST (Volume-limited
A-STars) imaging survey (De Rosa et al. \cite{DR11}, \cite{DR12}). It includes main
sequence B6 to A7-type stars within 75~pc, with projected separations $a \leq 100$~AU.
Imaging is done mainly in the K band with adaptive optics and medium to large telescopes
(3m Shane, 8m Gemini and VLT). Intermediate results show that among 148 multiple systems,
there are no more than 39\% binary, but at least 46\% triple and at least 15\% quadruple
ones. Tokovinin et al. \cite{T13} also detected close companions of nearby field stars.
A similar survey in the Sco OB2 association (Kouwenhoven et al. \cite{K05}, \cite{K07})
provided a CF of $40\pm4$\%, though some pollution by background stars or substellar objects
may be present. 

For chemically peculiar (CP) stars, Hubrig et al. \cite{H08} found MF$>50$\% among HgMn
stars with NACO at the VLT. Sch\"oller et al. \cite{S12} surveyed 28 roAp stars and found
6 probable companions with projected separation between 30 and 2400 AU. This confirms that
roAp stars are less often members of multiple systems than noAp ones, as already suspected
by Hubrig \cite{H00} from radial velocities. Balega et al. \cite{B11} surveyed 117
magnetic Ap and Bp stars with speckle interferometry at the 6m BTA telescope. They found
$MF=25$\%, with projected separations from 16 to 4290 AU.

Radial velocity (RV) surveys of normal A stars were performed e.g. by Abt \cite{A83},
who found 30--45\% of spectroscopic binaries (SB). Verschueren et al. \cite{V96} found
75\% of RV variables among 132 B-A stars in the Sco-Cen OB association. However, some of
them may be intrinsic variables, so the quoted figure should be considered as an upper
limit to the true frequency of multiple systems.

It is interesting to consider not only A and B stars proper, but also "retired" A-B stars,
i.e. red giants (RG) with the same masses, because their RVs are easier to determine. An
extensive, but as yet unexploited survey has been performed by Mermilliod et al. \cite{M07}
with the Coravel spectrovelocimeters at OHP and ESO La Silla. They observed RGs in open
clusters, found 264 SBs out of 1309 targets, and determined (adding 25 systems from the
literature) $MF=22$\%. They could determine the orbital parameters of 131 systems.
The advantage of this survey is that the age of each primary is known from the cluster age,
as well as its mass: the mere fact that the star lies in the clump allows a mass determination
to within $\pm 5$\% or so.
\begin{figure}[!t]
\begin{center}
 \includegraphics[width=5.0cm]{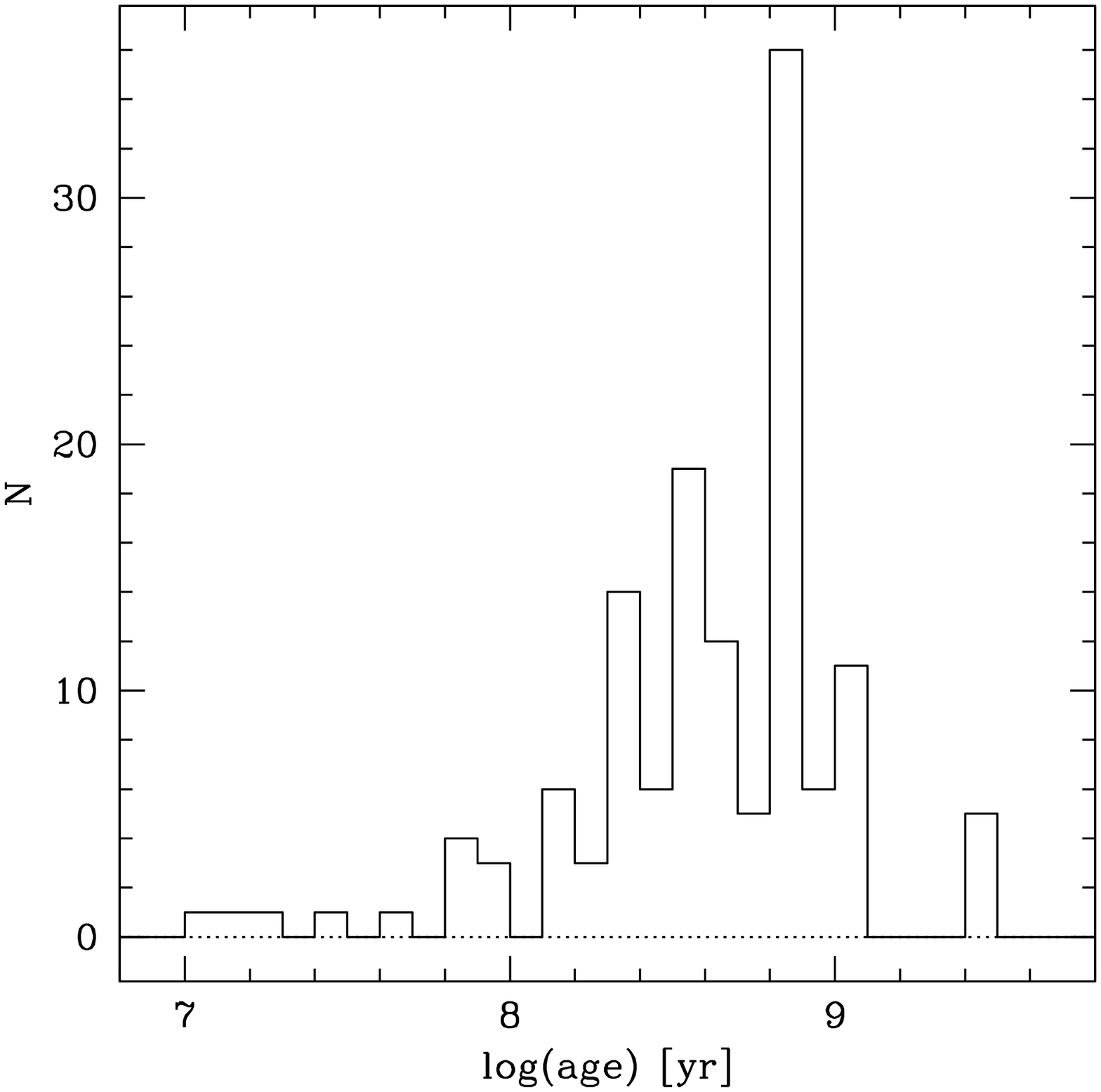}
 \includegraphics[width=5.0cm]{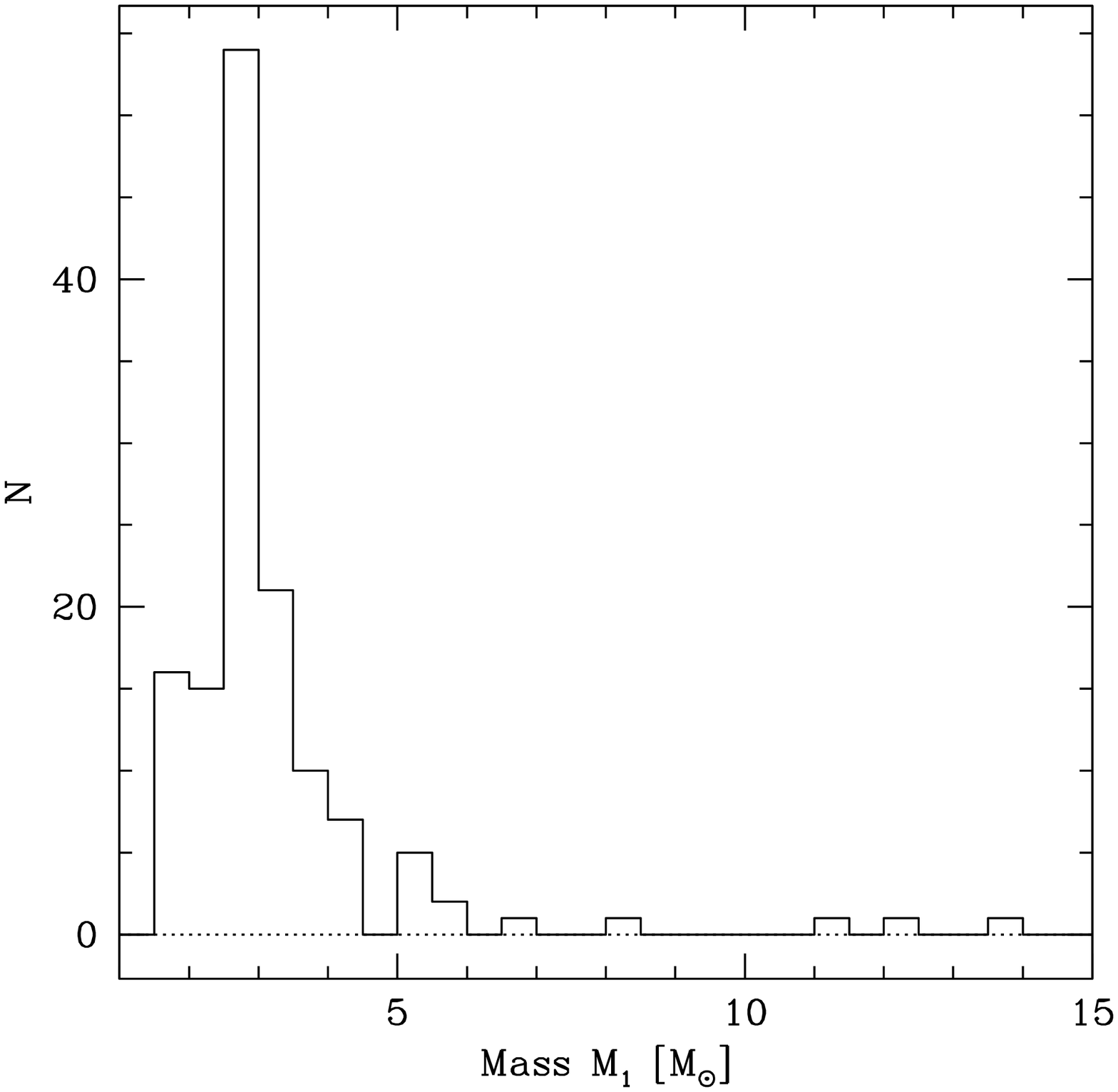}
\vspace{-5mm}
\caption[]{Left: Age distribution of the cluster red giants in binaries. Right: mass
distribution of the same.}
\label{age_mass_distr}
\end{center}
\end{figure}
The main caveat is that the present orbital parameters may differ
from the original ones because of dynamical evolution or mass exchange. Most binaries are
SB1s, so only the mass function is obtained. But, since $M_1$ is known, it is possible to
recover the mass ratio distribution in a statistical sense, assuming random inclinations.
We have simulated the mass function distribution by generating 10000 systems for each
observed primary, assuming a flat $q$ distributions (Fig.~\ref{fmdistr_noWD}). The overall shape of the
distribution is not recovered satisfactorily, because the peak at $\log(f(M))=-1.9$ is not
reproduced. The problem lies in the implicit assumption that the red giant is not only the
present primary, but has always been. Actually, some must have been secondaries in systems
where the former primary has since become a white dwarf (WD). With the very simplistic
assumption that the WD masses are distributed according to a Gaussian centred on 0.6~M$_\odot$
and with $\sigma=0.03$~M$_\odot$, the peak of the $\log[f(M)]$ distribution at $-1.9$ is
remarkably well reproduced, provided the relative number of WD companions is $\sim 23$\%
(Fig.~\ref{fmdistr_simple}). The relative number of such systems can be estimated assuming a
Salpeter IMF of the initial primaries, taking advantage of the narrow age and mass
distributions of the observed giants (Fig.~\ref{age_mass_distr}). The result, $\sim 25$\%, perfectly agrees
with the value needed in the simulation to fit the observed mass function distribution.
It does not depend on the $\gamma$ value, the index of the power law assumed
for the $q$ distribution ($f(q)\propto q^\gamma$): we adopted $\gamma=0$, but adopting e.g.
$-0.4$ would not change the position of the peak even though the overall fit would worsen.
Therefore, the result is robust.
\begin{figure}[!t]
\begin{center}
 \includegraphics[width=7.5cm]{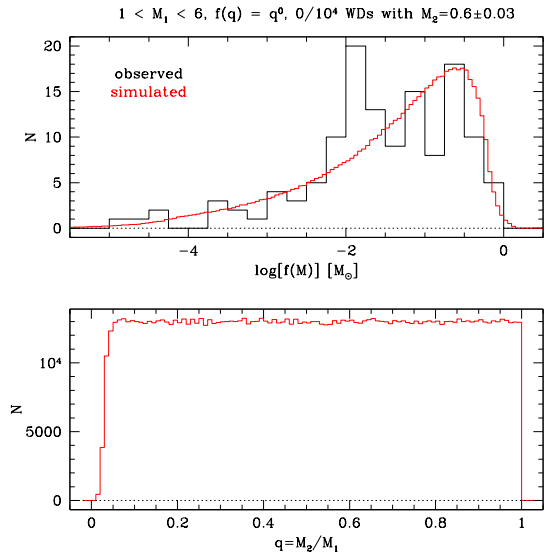}
\vspace{-5mm}
\caption[]{Top: Observed versus simulated mass function distribution, assuming the red giant
to be the most massive component in each binary system. No WD included. Bottom: mass ratio
distribution used in the simulation.}
\label{fmdistr_noWD}
\end{center}
\end{figure}
\begin{figure}[!t]
\begin{center}
 \includegraphics[width=7.5cm]{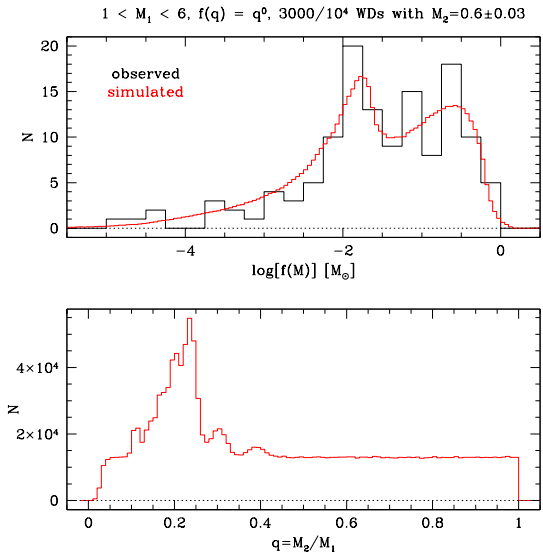}
\vspace{-5mm}
\caption[]{Top: Observed versus simulated mass function distribution, including WDs resulting
from former primaries. Bottom: Same as Fig.~\ref{fmdistr_noWD}, but including WDs with
$M=0.6\pm 0.03$~M$_\odot$.}
\label{fmdistr_simple}
\end{center}
\end{figure}

\begin{figure}[!t]
\begin{center}
 \includegraphics[width=7.5cm]{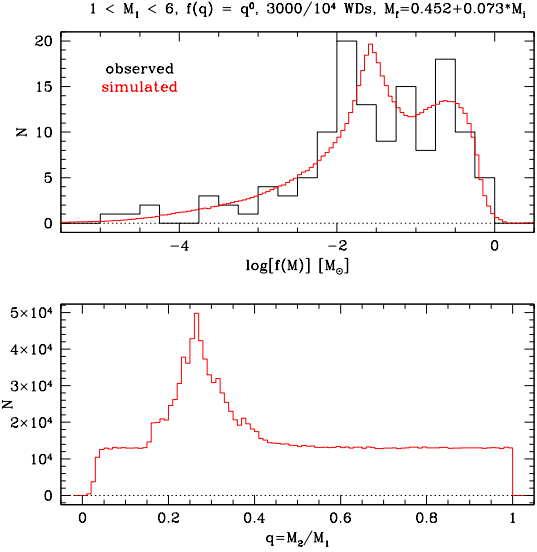}
\vspace{-5mm}
\caption[]{Top: Same as Fig.~\ref{fmdistr_simple}, but with WDs obeying the initial-final mass relation
of Zhao et al. \cite{Z12}. Bottom: Same as Fig.~~\ref{fmdistr_simple} but with WDs as in the top panel.}
\label{fmdistr_real}
\end{center}
\end{figure}

However, we have neglected the initial to final mass relation (IFMR) of intermediate mass
stars while attributing masses to WDs. The larger the progenitor mass, the larger the WD
mass (see e.g. Zhao et al. \cite{Z12}). Taking it into account, the fit becomes less satisfactory,
even though the simulation remains in qualitative agreement with the observations (Fig.~\ref{fmdistr_real}).
The reason for such a discrepancy remains unclear, but may be linked with binary interactions
like mass exchange. Indeed, we only considered here the case of components evolving in
isolation. Undoubtedly, a much more sophisticated simulation would be needed to interpret
the observations, and the latter should also be expanded to reach more significant
conclusions. Nevertheless, this approach represents a new way to look for WDs in open
clusters, and may prove useful to bring constraints on binary evolution. A moderate
observational effort would suffice to nearly double the number of systems with known
orbital parameters, thereby better characterizing the mass function distribution.

Regarding the CP stars, the first surveys were made by Abt \& Snowden \cite{AS73} on
47 magnetic Ap stars, and by Aikman \cite{A76} on 80 HgMn stars. They found MF$=20$\%
and $49$\% respectively, already suggesting a slightly low rate for the former and a high
one for the latter. The first review (Gerbaldi et al. \cite{G85}) concluded that MF is
lower for He-weak and Si stars than for normal stars, while it appeared normal for HgMn and
SrCrEu stars. The authors also noticed a significant lack of orbital periods shorter than
3 days for all Ap stars, including the HgMn ones, and an excess of high eccentricities.
They pointed out a lack of SB2s among magnetic Ap stars, while many exist among HgMn stars.
The next review (Carrier et al. \cite{C02}) confirmed the lack of orbital periods shorter
than 3 days (the apparent exception, HD 200405, was since proven to be an Am star, in
spite of being photometrically Ap). They showed that the apparent excess of large eccentricities
just results from the lack of short orbital periods, because the latter correspond to
zero eccentricities. Therefore, the relevant question is the lack of short periods, not
an excess of large eccentricities. Finally, they obtain a rather normal MF$\sim 44$\%.
An interesting, though little known survey of Am stars (Debernardi \cite{D02}) revealed
a very sharp limit at $P_\mathrm{orb}\sim 9$~days between small and large eccentricities. It also
suggests that, although a majority of Am stars are members of binary systems, they can
be divided in two populations according to the probable cause of their slow spin rotation: those
that are members of short period systems (with a peak at $P_\mathrm{orb}\sim 8$~days) and owe
their slow rotation to tidal effects, and those which are single or members of long
period systems and are slow rotators since their formation. Similar conclusions were
reached by Carquillat \& Prieur \cite{CP07}.

By photometric surveys, I mean the detection of eclipsing binary (EB) systems. The OGLE survey
(e.g. Udalski et al. \cite{U98}) is the most famous example. Mazeh et al. \cite{M06}
selected $\sim 900$ EBs among 2600 detected in the LMC by Wyrzykowski et al. \cite{W03} and
estimated the detection probability of each object, after having interpreted the lightcurves
which give the orbital inclination and the radii in units of the orbital semi-major axis.
Correcting for the detection probability, they found that for $P_\mathrm{orb}<10$~days, the
$\log(P_\mathrm{orb})$ distribution is flat. Whether this result is correct is yet to be verified
in the light of more reliable detection methods used since.

Some individual objects are worth of special attention: AO Vel is a quadruple system made
of two close binaries, one of which is an eclipsing Si star and brakes the $P > 3$~days rule
with $P_\mathrm{orb}=1.58$~days. The other system includes two HgMn stars (Gonzalez et al. \cite{G10}
and references therein). HD 123335 is a long period, highly eccentric SB2 EB with only
one eclipse, hosting two 5 M$_\odot$ stars of the He-weak SrTi type (Hensberge et al. \cite{H04}).
The primary rotates more slowly than the pseudosynchronous velocity. AR Aur is an EB with
an HgMn primary and a probably Am secondary (Folsom et al. \cite{F10}). It is the very first
HgMn star where intrinsic spectral variations have been suspected (Takeda \& Takada \cite{TT79}).
The A0pSi SB1 system $\tau^9$ Eri is not an EB but has two photometric periods, the shorter of which
is the magnetic one (Manfroid et al. \cite{M85}). The origin of the longer period remains unknown: it differs
from the orbital period, so it might be due to a magnetic secondary, though the amplitude
appears exeptionally large under that assumption (North \& Debernardi \cite{ND04}).

\section{The glorious fate of the happy few: the SNe Ia}
Some systems hosting A-type stars will end as Type Ia supernovae, because the latter consist
in the thermonuclear explosion of a carbon-oxygen WD, the remnant of an intermediate mass star. Which kind
and which fraction of binaries with an A-type primary will undergo such a fate?
The answer is not easy, because such systems must pass through a poorly understood  common
envelope (CE) phase before ending as SNe Ia, and there are at least two possible progenitors:
the double degenerate (DD) ones, systems hosting two WDs, and the single degenerate (SD) ones
hosting an accreting WD and a non-degenerate donor (e.g. Maoz \& Mannucci \cite{MM12}). Ivanova et al.
\cite{I06} showed in their Fig.~1 the regions in the $\log P$ vs $M_\mathrm{p}$ (where $P$ is the orbital
period and $M_\mathrm{p}$ the initial mass of the primary) where circular binary
systems can be progenitors of SD systems, or cataclysmic variables (CV) from the observers'
point of view, some of which in turn are
progenitors of SNe Ia. (see e.g. Hachisu et al. \cite{HKN08}). Ivanova et al.
obtained their Fig.~1 by evolving 100'000 systems with $0.5<M_\mathrm{p}<10$, assuming a flat $q$
distribution and a flat $\log P$ distribution between $P=1$ and $10\,000$ days. Their Fig.~2
shows the same for progenitors of AM CVn systems, which generally have passed through two CE phases
before becoming ultra-short period systems with a degenerate donor (a low mass helium star)
and a WD accretor. AM CVn systems are, therefore, good progenitor candidates for SNe Ia events
through the DD channel. For both channels, the ultimate fate of the system depends on the mass
of the WD: it will explode as a SN Ia only of it is made of CO, which restricts its mass to
the range $\sim 0.8-1.1$~M$_\odot$. The lower mass limit is defined by the possibility to trigger
carbon burning, while the upper limit coincides with the transition from CO to ONeMg composition
(Geier et al. \cite{G13}). A $0.8$~M$_\odot$ WD is the remnant of a $3-4$~M$_\odot$ main
sequence star, i.e. a mid- to late-B type star.

A clever, though less popular path to SNe Ia is the core-degenerate (CD) scenario:
in systems hosting at least one component with $M > 2.3$~M$_\odot$ and wide enough that the CE phase
occurs on the AGB but not on the RGB, the WD resulting from the former primary can merge
with the hot degenerate core of the AGB (evolved initial secondary) at the end of, or shortly after
the CE phase (Ilkov \& Soker \cite{IS12}, \cite{IS13}). The result is a degenerate object with a mass larger
than the Chandrasekhar limit ($M > M_\mathrm{Ch}(V_\mathrm{rot}=0)$), but this object is nevertheless stable thanks to rapid
rotation. Then, it is spun down through magnetodipole radiation and explodes as soon as its
rotation velocity is low enough that its mass exceeds $M_\mathrm{Ch}(V_\mathrm{rot})$ {\sl for the
current rotation velocity $V_\mathrm{rot}$}. For a rigidly and rapidly rotating WD, $M_\mathrm{Ch}\approx 1.48$~M$_\odot$
(Yoon \& Langer \cite{YL04}), so the CD scenario predicts a mass range $1.40-1.48$~M$_\odot$
for the SN Ia progenitors in cases of long delay times (for shorter
delays, the mass may exceed $1.5$). The delay time can be as short as a Myr or as long as 10 Gyr.

\section{Promise of the dawn: Gaia and other future surveys}
In the quest for reliable, unbiased samples on which frequency and properties of multiple
systems can be anchored, future large surveys will certainly allow dramatic progress.
In particular, the Gaia space mission will provide very large volume-limited samples of both
astrometric and spectroscopic binaries. About 30 million of the former are expected, and 8
million of the latter, 59\% of which should be SB2. On top of that, 4 million EBs should be
detected, 12\% of which will also be detected as SB1 or SB2 (Eyer et al. \cite{E13}). The detection probability of the
EBs, however, will be more difficult to determine than that of the SBs, especially for long
period objects. Gaia will also provide parallaxes for many EBs that are already
known from e.g. the OGLE bulge survey, and for those to be discovered by LSST. However,
LSST will measure objects with $V\geq 16$, for which the astrometric precision will not be
optimal, so the interest will be limited to nearby EBs hosting intrinsically faint dwarfs. Long period
systems, especially those detected through astrometry, will have to be followed up from the
ground, which will require dedicated long term programs. LSST should prove very efficient
in discovering and following long period EBs photometrically; spectroscopic follow-up of these
objects will require a considerable effort.

\bigskip
{\it Acknowledgements.} I thank Dr. Laurent Eyer for some details about Gaia and comments
on the last Section.


\begin{thebibliography}{99}

\bibitem{A83}
{Abt H.A.} 1983, ARA\&A, 21, 343

\bibitem{AS73}
{Abt H.A., Snowden M.S.} 1973, ApJS, 25, 137

\bibitem{A76}
{Aikman G.C.L.} 1976, PDAO, 14, 379

\bibitem{B11}
{Balega Y.Y.} 2011, AN, 332, 978

\bibitem{CP07}
{Carquillat J.-M., Prieur J.-L.} 2007, MNRAS, 380, 1064

\bibitem{C02}
{Carrier F., North P., Udry S. et al.} 2002, \aap, 394, 151

\bibitem{D02}
{Debernardi Y.} 2002, PhD thesis, Geneva University

\bibitem{DR11}
{De Rosa R. J., Bulger J., Patience J. et al.} 2011, MNRAS, 415, 854

\bibitem{DR12}
{De Rosa R. J., Patience J., Vigan A. et al.} 2012, MNRAS, 422, 2765

\bibitem{DK13}
{Duch\^ene G., Kraus A.} 2013, ARA\&A, 51, 269

\bibitem{E13}
{Eyer L., Holl B., Pourbaix D. et al.} 2013, CEAB, 37, 115

\bibitem{F10}
{Folsom C.P., Kochukhov, O., Wade, G. A. et al.} 2010, MNRAS, 407, 2383

\bibitem{G13}
{Geier S., Marsh T.R., Wang B. et al.} 2013, \aap, 554, A54

\bibitem{G85}
{Gerbaldi M., Floquet M., Hauck B.} 1985, \aap, 146, 341

\bibitem{G10}
{Gonzalez J.F., Hubrig S., Castelli F.} 2010, MNRAS, 402, 2539

\bibitem{HKN08}
{Hachisu I., Kato M., Nomoto K.} 2008, \apj 679, 1390

\bibitem{H04}
{Hensberge H., Nitschelm C., Freyhammer L.M. et al.} 2004, ASPC, 318, 309

\bibitem{H00}
{Hubrig S., Karchenko N., Mathys G. et al.} 2000, \aap, 355, 1031

\bibitem{H08}
{Hubrig S., Ageorges N., Sch\"{o}ller M.} 2008, in Multiple stars across the
HR diagram, ESO Astrophysics Symposia, Springer, Berlin, Heidelberg, p. 155

\bibitem{I06}
{Ivanova N., Heinke C. O., Rasio F. A., et al.} 2006, MNRAS, 372, 1043

\bibitem{IS12}
{Ilkov M., Soker N.} 2012, MNRAS, 419, 1695

\bibitem{IS13}
{Ilkov M., Soker N.} 2013, MNRAS, 428, 579

\bibitem{K05}
{Kouwenhoven M.B.N., Brown A.G.A., Zinnecker H. et al.} 2005, \aap, 430, 137

\bibitem{K07}
{Kouwenhoven M.B.N., Brown A.G.A., Kaper L.} 2007, \aap, 464, 581

\bibitem{M85}
{Manfroid J., Mathys G., Heck A.} 1985, \aap, 144, 251

\bibitem{MM12}
{Maoz D., Mannucci F.} 2012, PASA, 29, 447

\bibitem{M06}
{Mazeh T., Tamuz O., North P.} 2006, MNRAS, 367, 1531

\bibitem{M07}
{Mermilliod J.-C., Andersen J., Latham D.W., Mayor M.} 2007, \aap, 473, 829

\bibitem{ND04}
{North P., Debernardi Y.} 2004, ASPC, 318, 297

\bibitem{S12}
{Sch\"{o}ller M., Correia S., Hubrig S.} 2012, \aap, 545, A38

\bibitem{TT79}
{Takeda Y., Takada M.} 1979, PASJ, 31, 821

\bibitem{T97}
{Tokovinin A.} 1997, \aaps, 124, 75

\bibitem{T13}
{Tokovinin A., Hartung M., Hayward T.L.} 2013, AJ, 146, 8

\bibitem{U98}
{Udalski A., Soszynski I., Szymanski M. et al.} 1998, \actaa, 48, 563

\bibitem{V96}
{Verschueren W., David M., Brown A.G.A.} 1996, ASPC, 90, 131

\bibitem{W03}
{Wyrzykowski L., Udalski A., Kubiak M. et al.} 2003, \actaa, 53, 1


\bibitem{YL04}
{Yoon S.-C., Langer N.} 2004, A\&A, 419, 623

\bibitem{Z12}
{Zhao J.K., Oswalt T.D., Willson L.A. et al.} 2012, ApJ, 746, 144

\end{thebibliography}
\end{document}